\documentclass{WileyMSP-template}

\usepackage{xcolor}
\usepackage{booktabs}
\usepackage{siunitx}
\usepackage[T1]{fontenc}
\DeclareSIUnit{\gauss}{G}
\DeclareSIUnit{\oersted}{Oe}

\begin{document}

%\pagestyle{fancy}
%\rhead{\includegraphics[width=2.5cm]{vch-logo.png}}

\title{Engineering a Correlated Narrow-Gap Semiconductor: Effects of Ga Substitution in EuZn$_2$P$_2$}

\maketitle

\author{Mateus Dutra}
\author{Esteban Marulanda}
\author{Gustavo G. Vasques}
\author{Jaime Ferreira Oliveira}
\author{Pedro C. Sabino}
\author{Rebeca Brigato Delgado}
\author{Leticie Mendonça-Fereira}
\author{Adriano R. V. Benvenho}
\author{Elisa Baggio-Saitovitch}
\author{Ricardo K. Machado}
\author{Nicolas M. Kawahala}
\author{Julian Munevar*}
\author{Marcos A. Avila*}
\author{Felix G. G. Hernandez*}

\begin{affiliations}
M. Dutra, G. G. Vasques, P. C. Sabino, R. B. Delgado, L. Mendonça-Fereira, A. R. V. Benvenho, M. A. Avila\\
CCNH, Universidade Federal do ABC (UFABC), 09210-580, Santo Andre, SP, Brazil\\
Email Address: avila@ufabc.edu.br\\

E. Marulanda, R. K. Machado, N. M. Kawahala, F. G. G. Hernandez\\
Instituto de Física, Universidade de São Paulo (USP), 05508-090, São Paulo, SP, Brazil\\
Email Address: felixggh@if.usp.br

E. Baggio-Saitovitch, J. F. Oliveira\\
Centro Brasileiro de Pesquisas Físicas, Rio de Janeiro, RJ, 2290-180 Brazil\\

J. Munevar\\
Departamento de Física, Universidad del Valle, Cali, A. A. 25360 Colombia\\
Email Address: julian.munevar@correounivalle.edu.co

\end{affiliations}

% Keywords: Please provide a minimum of three and a maximum of seven keywords, separated by commas


% Abstract should be written in the present tense and impersonal style (i.e., avoid we), and be at most 200 words long

\begin{abstract}
The effect of Ga substitution on the electronic, magnetic, and low-energy responses of the Zintl phase EuZn$_2$P$_2$ is investigated by electrical transport, electron spin resonance (ESR), and terahertz time-domain spectroscopy (THz-TDS). Incorporating Ga into EuZn$_2$P$_2$ (EuZn$_{1.8}$Ga$_{0.2}$P$_2$) reduces the electrical resistivity, indicating enhanced free-carrier density and a narrowed semiconducting gap. ESR confirms the persistence of Eu$^{2+}$ moments while showing a crossover from a Lorentzian to a Dysonian lineshape, consistent with reduced skin depth, increased carrier density, and the emergence of diffusive contributions. Ga-substituted compound display pronounced negative magnetoresistance linked to magnetic-polaron formation. THz-TDS reveals strong low-frequency absorption and a notable enhancement of the Drude conductivity in the substituted material, together with an increased carrier scattering time and enhanced carrier-density--to--effective-mass ratio. These results demonstrate that Ga substitution tunes charge transport, carrier dynamics, and short-range magnetic correlations in EuZn$_2$P$_2$, establishing EuZn$_{1.8}$Ga$_{0.2}$P$_2$ as a promising platform for engineering correlated narrow-gap magnetic semiconductors with enhanced electronic and spin-dependent functionalities.
\end{abstract}

% Text: Please use section headings and subheadings as specifoied below. For communications, all section headings apart from Experimental Section should be removed
% Please make the first reference to a display item bold: \textbf{Figure 1}
% Do not abbreviate Figure, Equation, etc.; display items are always singular, i.e., Figure 1 and 2.
% Equations are always singular, i.e., Equation 1 and 2, and should be inserted using the {equation} environment, not as graphics
% Please do not use footnotes in the text, additional information can be added to the Reference list.

\section{Introduction}

Band-structure engineering provides a powerful route for tailoring the electronic, magnetic, and optical responses of solid-state materials. A classical strategy for achieving such control is chemical substitution, which can modulate carrier density, lattice structure, and electronic correlations. Zintl phases are a prominent class of semiconductors in this context: their stability arises from electron-counting rules that enforce charge balance, resulting in predominantly insulating or semiconducting behavior \cite{shuai2017recent, freer2022key, westbrook1995intermetallic}. Their tunable bonding chemistry and structural diversity make them appealing platforms for designing functional materials with tailored properties.

Among Zintl phases, Eu-based compounds crystallizing in the CaAl$_2$Si$_2$-type structure (space group $P\overline{3}m1$) have recently attracted significant interest. Their layered arrangement---Eu$^{2+}$ cations separated by [M$_2$X$_2$]$^{2-}$ frameworks---naturally decouples magnetic and conducting units, allowing magnetic and electronic degrees of freedom to be tuned with unusual flexibility. This has led to reports of diverse emergent phenomena including colossal magnetoresistance, low-temperature magnetic orderings, optoelectronic responses, and even possible non-trivial electronic band states, though their existence is yet to be confirmed \cite{wang2021colossal, ma2019spin, du2022consecutive, luo2023colossal}. Such rich behavior underscores their potential for applications in spintronic, optoelectronic, and magnonic technologies \cite{berry2022type, berry2021antiferro}.

EuZn$_2$P$_2$ exemplifies the sensitivity of these materials to subtle chemical and structural variations. It is a narrow-gap semiconductor that exhibits antiferromagnetic (AFM) order near \qty{23.5}{\K}, colossal magnetoresistance around \qty{150}{\K}, and strong semiconducting behavior \cite{berry2022type, krebber2023colossal}. Recent studies have demonstrated that its ground state depends strongly on synthesis conditions: samples grown via Sn or In metallic flux remain AFM semiconductors, whereas their growth using NaCl/KCl flux produces a metallic ferromagnet, due to the formation of Eu vacancies \cite{chen2024carrier}. Despite this sensitivity, no systematic investigation has yet examined how intentional chemical substitutions modify the physical properties of EuZn$_2$P$_2$.

Understanding such effects is particularly relevant because EuZn$_2$P$_2$ has an experimentally estimated band gap between \num{100} and \qty{350}{\meV} \cite{berry2022type, krebber2023colossal, chen2024carrier, singh2023superexchange, PhysRevB.110.014421}, anticipating it in a technologically important regime. A recent study reported that EuZn$_2$P$_2$ exhibits a photovoltaic-like photoresponse under Pt contact, generating photocurrent even at zero bias due to a light-induced photovoltage \cite{dutra2025spin}. However, impurities and defect states within the gap can degrade their performance by introducing recombination centers and suppressing the optical sensitivity \cite{chen2023carrier}. Identifying substitutional strategies that tune the band gap while preserving a clean electronic structure is therefore critical for developing broadband photodetectors and related devices \cite{lian2024review, long2019progress}.

Narrow-gap semiconductors in this energy window are also central to emerging solid-state approaches for probing dark matter. Dark matter, though non-luminous, exerts gravitational influence and remains one of the outstanding mysteries in modern physics \cite{garrett2011dark}. Leading candidates include weakly interacting massive particles (WIMPs) and axions \cite{bertone2018new}. While WIMPs are heavy (\qty{2}{\GeV}--\qty{100}{\TeV}) and interact only via weak and gravitational forces \cite{roszkowski2018wimp}, axions can convert into photons in the presence of magnetic fields \cite{sikivie1983experimental, sikivie2021invisible}. Experiments such as SPLENDOR (Search for Particles of Light Dark Matter with Narrow-gap Semiconductors) specifically exploit the low-energy excitation thresholds of materials like EuZn$_2$P$_2$ to search for such hypothetical particles \cite{abbamonte2025splendor}. For example, Eu$_5$In$_2$Sb$_6$, with a $\sim$\qty{50}{\meV} gap, is currently under investigation as a detector material. Thus, chemical substitutions must be implemented carefully, as they can introduce defect centers detrimental to both photon detection and electronic transport.

Motivated by these considerations, this work investigates Ga substitution in EuZn$_2$P$_2$ single crystals grown by the Sn flux method, that nominally results in  EuZn$_{1.8}$Ga$_{0.2}$P$_2$, as a strategy to reduce the band gap while preserving a clean energy gap. Through temperature-dependent resistivity, magnetoresistance, electron spin resonance, and terahertz time-domain spectroscopy, we elucidate how Ga incorporation modifies the interplay between magnetic order, charge transport, and low-energy electrodynamics. These results demonstrate that Ga substitution offers a promising route for tuning EuZn$_2$P$_2$ toward applications spanning semiconducting electronics, spintronics, optoelectronics, and quantum sensing.

\section{Results and Discussion}

\subsection{Crystal Structure}

\begin{figure}[htpb]
   \centering
   \includegraphics[scale=.36]{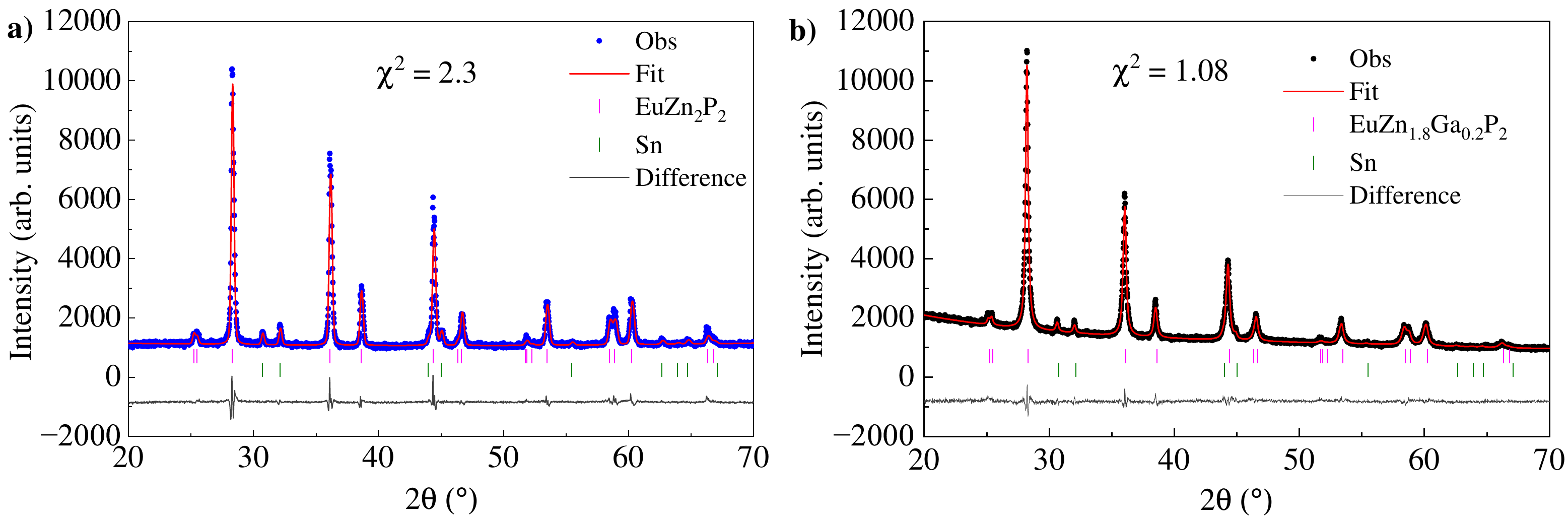}
  \caption{powder X-ray diffraction pattern of crushed single crystals of a) EuZn$_2$P$_2$ and b) EuZn$_{1.8}$Ga$_{0.2}$P$_2$. The experimental data are presented in blue and black for EuZn$_2$P$_2$ and EuZn$_{1.8}$Ga$_{0.2}$P$_2$, respectively. The Rietveld refinement is shown in red, the difference between experiment and model is shown in gray, and the Bragg reflections corresponding to EuZn$_2$P$_2$, EuZn$_{1.8}$Ga$_{0.2}$P$_2$ and Sn flux are shown as vertical lines.} 
   \label{fig:XRD}
\end{figure}

The powder X-ray diffraction (PXRD) pattern of EuZn$_{2}$P$_2$ and EuZn$_{1.8}$Ga$_{0.2}$P$_2$ are shown in \textbf{Figure~\ref{fig:XRD}}. The Rietveld refinement confirms the trigonal structure with the $P\overline{3}m1$ space group plus small peaks coming from the residual Sn flux. The lattice parameters of EuZn$_2$P$_2$ are $a=\qty{4.0850(3)}{\angstrom}$, $c=\qty{7.0029(3)}{\angstrom}$, $V=\qty{101.0(8)}{\cubic\angstrom}$, whereas for the Ga-substituted compound the extracted parameters increased to $a=\qty{4.0944(2)}{\angstrom}$,  $c=\qty{7.0213(7)}{\angstrom}$, $V=\qty{101.9(3)}{\cubic\angstrom}$; the unit cell volume of Ga-substituted compound is near \qty{1}{\percent} larger compared with the pristine compound. The small deviation of the peaks, reflects the effect of the crystal lattice by Ga$^{3+}$ substitution at Zn sites. Although Ga$^{3+}$ has a smaller ionic radius than Zn$^{2+}$ in tetrahedral coordination (\num{0.62} vs. \qty{0.74}{\pm}), its substitution in the lattice is expected to inject additional carriers into the [MP$_4$] (M = transition metal) tetrahedron. This carrier doping can increase the electronic density within the framework, potentially driving an accommodation of the crystal structure. Another possibility is that the lattice expansion arises from enhanced ionic repulsion within the [Zn$_2$P$_2$]$^{2-}$ framework upon Ga substitution. 
In this scenario, it is plausible that electronic reorganization plays a more significant role than the nominal ionic radius difference.

Generally, when attempting to perform substitutions to move from one compound to another along a solid solution series, it is common to encounter difficulties in the intermediate region of substitutions. The case of the EuZn$_2$P$_2$ system with Zn substituted by Ga, it is expected that at \qty{100}{\percent} Ga the end member would be EuGa$_2$P$_2$. However, in many instances, it is not possible to achieve a complete substitution range due to factors such as solubility limits, structural instabilities, or phase transitions. EuGa$_2$P$_2$ has been reported to retain a Zintl-like character \cite{goforth2009magnetism}, yet it exhibits metallic behavior, an intriguing feature, since charge balance is typically required for the stabilization of Zintl phases. This compound crystallizes in a monoclinic structure ($P2/m$ space group), and excess Ga content during synthesis can promote the formation of impurity phases, both because of the high synthesis temperatures and the strong tendency of rare-earth elements to form binary compounds, especially with Ga. In our case, by maintaining a moderate Ga substitution level, we ensured the absence of secondary phases, as confirmed by the PXRD, thus providing greater confidence in the results discussed below.

\subsection{Electrical Resistivity and Magnetoresistance}

\begin{figure}[htpb]
   \centering
   \includegraphics[scale=.43]{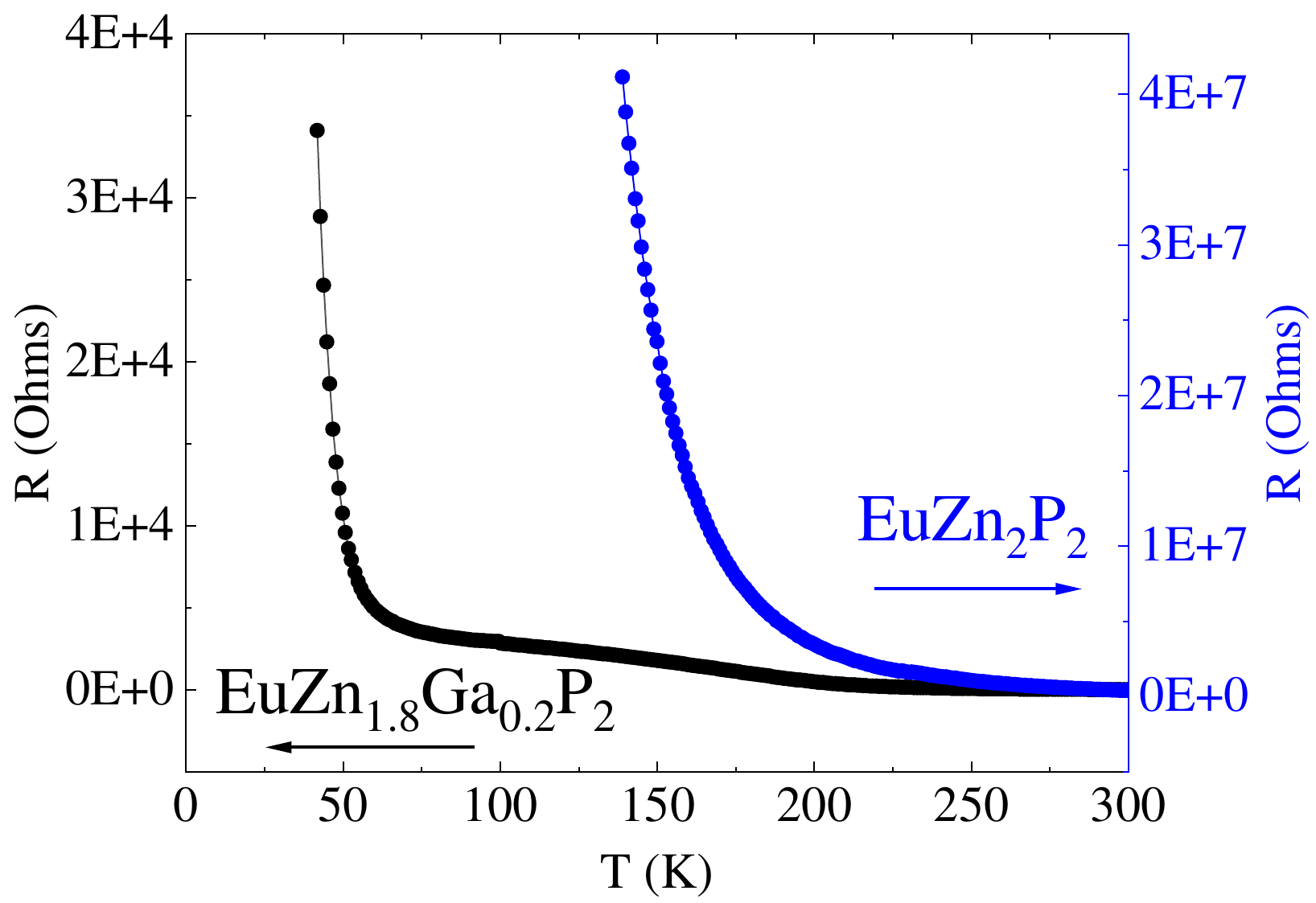}
  \caption{Comparison of the electrical resistances of EuZn$_2$P$_2$ (blue) and EuZn$_{1.8}$Ga$_{0.2}$P$_2$ (black). The activated regime in EuZn$_2$P$_2$ begins around 150 K, whereas in EuZn$_{1.8}$Ga$_{0.2}$P$_2$ it starts near 50 K.} 
   \label{fig:resistance}
\end{figure}

In \textbf{Figure~\ref{fig:resistance}} we present the electrical resistance data measured for EuZn$_2$P$_2$ (blue) and for EuZn$_{1.8}$Ga$_{0.2}$P$_2$ (black). We observe clear shift of the resistance upturn towards lower temperatures, especially in the activated regime.

\begin{figure}[htpb]
   \centering
   \includegraphics[scale=.34]{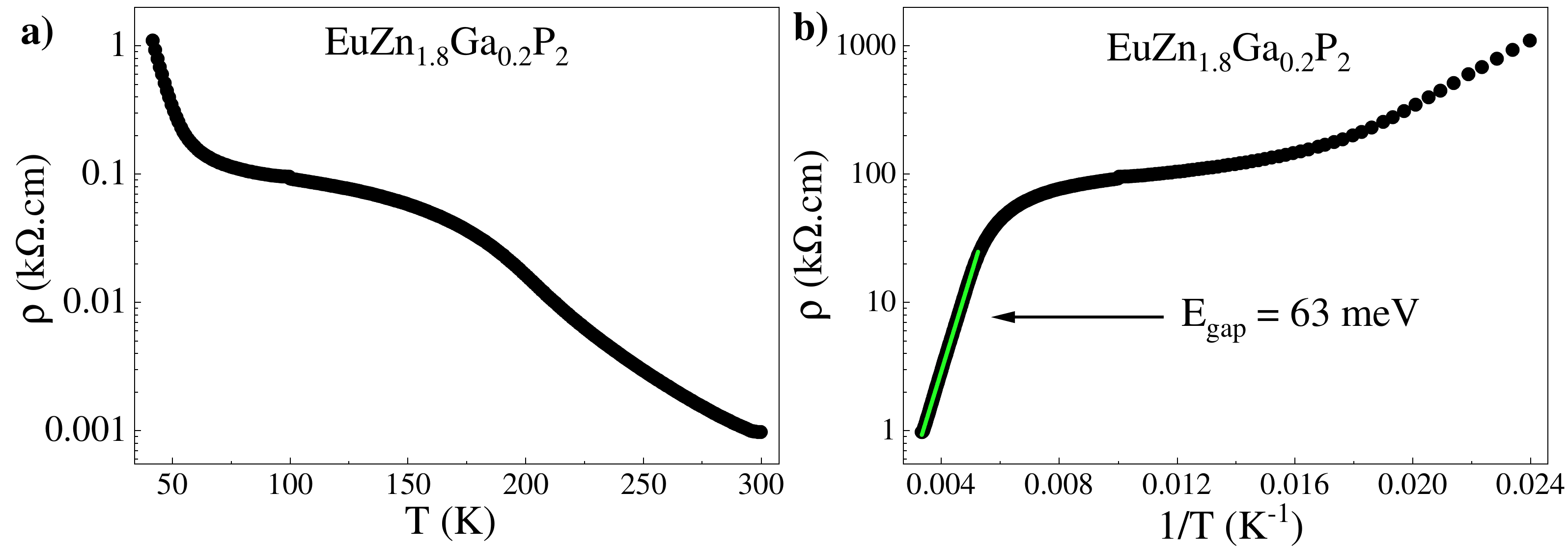}
  \caption{a) Electrical resistivity of EuZn$_{1.8}$Ga$_{0.2}$P$_2$ from \num{45} to \qty{300}{\K}. At least three distinct transport regimes can be identified in this temperature range. b) Arrhenius plot with linear fitting, yielding an estimated energy gap of approximately \qty{63}{\meV}.} 
   \label{fig:resistivity}
\end{figure}

To quantitatively analyze the modifications in the temperature-dependent electrical resistivity scale, we examine the data shown in \textbf{Figure~\ref{fig:resistivity}a}. The temperature-dependent resistivity $\rho(T)$ of the Ga-substituted compound shows a reduction in the electrical resistivity of approximately two orders of magnitude at \qty{150}{\K} relative to the electrical resistivity of the pristine compound reported in previous works \cite{berry2022type, singh2023superexchange, PhysRevB.110.014421, chen2023carrier}. The activation behavior characteristic of semiconductors is observed, and we estimate an energy gap of $E_{gap}=63$~meV for the Ga-substituted compound, a reduction of $\sim\qty{45}{\percent}$, as shown by the Arrhenius plot in Figure~\ref{fig:resistivity}b. There are at least three distinct regimes in the resistivity data in Figure~\ref{fig:resistivity}, and upon cooling, a new characteristic energy scale emerges between \qty{70}{\K} and \qty{170}{\K}. This behavior is consistent with the existence of incipient magnetic fluctuations of which we observe further evidence from the MR data. 

\begin{figure}[htpb]
   \centering
   \includegraphics[scale=0.34]{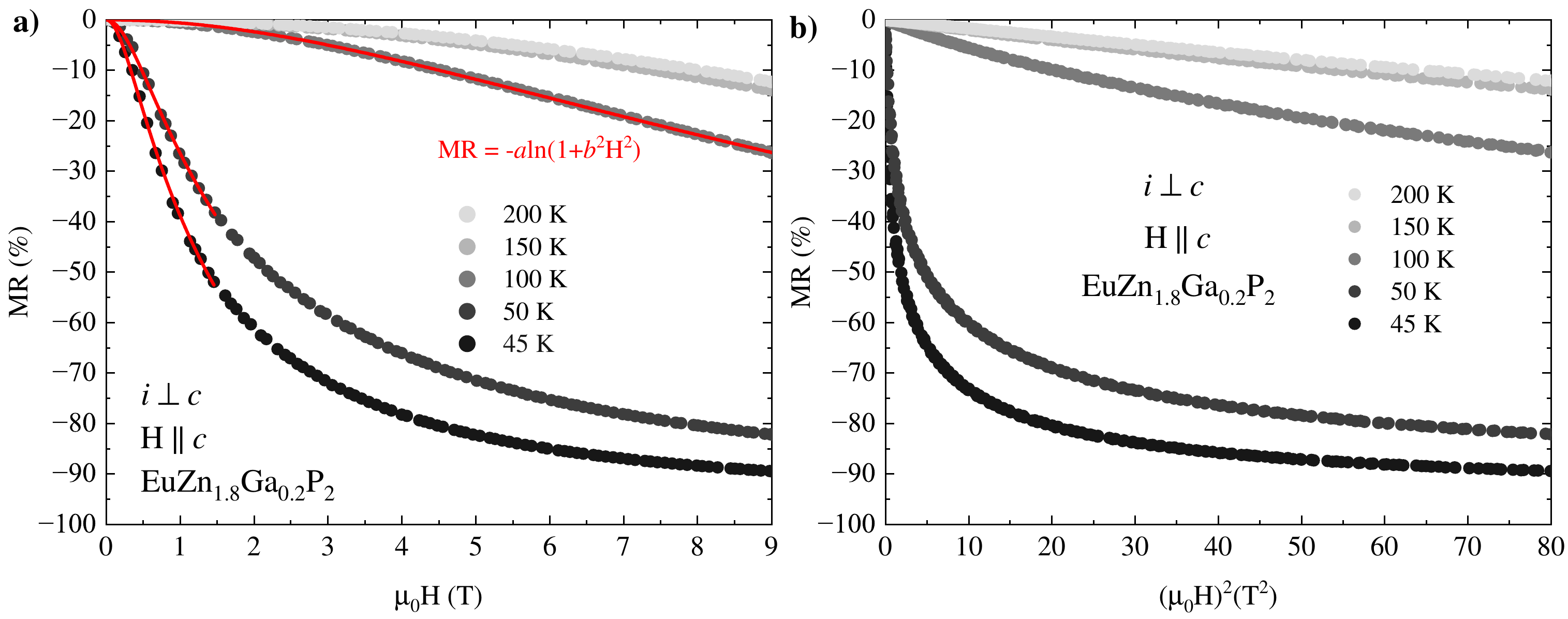}
  \caption{Isothermal magnetoresistance of EuZn$_{1.8}$Ga$_{0.2}$P$_2$ with $i \perp c$ and $H \parallel c$ at various temperatures. a) MR as a function of applied magnetic field from \num{0} to \qty{9}{\T}. b) MR as a function of $(\mu_0 H)^2$.} 
   \label{fig:MR}
\end{figure}

The \textbf{Figure~\ref{fig:MR}a} show the negative magnetoresistance data for EuZn$_{1.8}$Ga$_{0.2}$P$_2$. At \qty{45}{\K} the MR reaches $\sim\qty{90}{\percent}$ at \qty{9}{\T}, using:

\begin{equation}
    MR=\frac{R(H)-R(0)}{R(0)} \times \qty{100}{\percent},
\end{equation}

where $R(H)$ and $R(0)$ are the electrical resistance in the presence and absence of a magnetic field, respectively. 
From Figure~\ref{fig:MR}b, where the magnetoresistance is plotted as a function of $B^2$, we observe that the isothermal curves at 150 and 100~K follow a classical quadratic behavior. In contrast, below 100~K a small but clear deviation from the $B^2$ dependence emerges, indicating that additional scattering mechanisms start to contribute to the electronic transport.

Several distinct features can be observed when comparing the MR for pristine reported in previous works \cite{krebber2023colossal, cook2025magnetic} and the Ga-substituted compound. The abrupt increase in resistivity observed in the pristine compound around \qty{150}{\K} may be directly related to magnetic-polaron formation, as proposed in previous studies \cite{cook2025magnetic}. Magnetic polarons---quasiparticles formed by Eu$^{2+}$ spins---can localize charge carriers, leading to a sudden rise in resistivity, precisely in the temperature range where colossal magnetoresistance (CMR) is observed in the pristine compound. This behavior highlights the presence of strong magnetic and electronic correlations in EuZn$_2$P$_2$.

The CMR in the Ga-substituted compound remains remarkably high (\qty{90}{\percent}), indicating that even with additional carriers, magnetic-polaron mechanisms remain active in EuZn$_{1.8}$Ga$_{0.2}$P$_2$ (Figure~\ref{fig:MR}a). Maintaining large CMR while increasing carrier concentration at low substitution levels is potentially beneficial in a technological perspective, although larger substitution levels might suppress the electronic correlations and magnetic fluctuations responsible for the CMR effect.
At \qty{45}{\K} and \qty{50}{\K}, the MR decreases by approximately \qty{40}{\percent} in EuZn$_{1.8}$Ga$_{0.2}$P$_2$ under \qty{1}{\T}, following $n=1.5$ in the power law $MR = AH^n$, where $A$ is a constant and $H$ is the magnetic field, consistent with the semi-empirical Khosla--Fischer model \cite{MR_localized_1970} as shown in Figure~\ref{fig:MR}a:

\begin{equation}
    \frac{\Delta\rho}{\rho(0)} = -aln(1+b^2H^2)
\end{equation}

\noindent where $a$ is related to the exchange parameter and average magnetization of the system, and $b$ is related to temperature and exchange parameters given by the following expressions:

\begin{equation}
    a=A_1J\eta(E_F)[S(S+1)+\langle M^2\rangle],
\end{equation}

\noindent where

\begin{equation}
    A_1=AN_A\frac{\sigma_J^2}{\sigma_0^2},
\end{equation}

\noindent and

\begin{equation}
    b^2=\left[1+4S^2\pi^2\left(\frac{2J\eta(E_F)}{g_0}\right)^4\right]\frac{g_0^2\mu_0^2}{(\alpha kT)^2},
\end{equation}

\noindent where $\eta(E_F)$ is the density of states at the Fermi level, $\langle M \rangle$ is the average magnetization, $A$ is a constant associated with exchange interactions, $\sigma_0^2$ and $\sigma_J^2$ are the scattering and exchange scattering cross sections, respectively, and $N_A$ is Avogadro’s number. $J$ is the exchange constant, $g_0$ is the unshifted $g$-factor, and $\alpha$ is a phenomenological parameter of the model.
As expected from the model, $a$ increases with temperature (from \num{0.18} at \qty{45}{\K} to \num{0.31} at \qty{100}{\K}), while $b$ decreases due to its $1/T$ dependence, ranging from \qty{1.61}{\per\oersted} at \qty{45}{\K} to \qty{0.37}{\per\oersted} at \qty{100}{\K}.

The Khosla--Fischer expression describes negative MR in magnetic semiconductors at weak magnetic fields, where localized magnetic moments scatter conduction electrons. When a magnetic field is applied, the moments align, reducing spin-disorder (magnetic disorder arises from the random orientation of the Eu$^{2+}$ moments) scattering and thus producing negative MR. The fitted region in Figure~\ref{fig:MR}a is consistent with short-range magnetic interactions and the Khosla--Fischer interpretation. As temperature increases, the influence of short-range magnetic interactions is reduced, and the materials' MR becomes quadratic, as shown in Figure~\ref{fig:MR}b. This change correlates with the three transport regimes observed in Figure~\ref{fig:resistivity} and indicates that they are strongly affected by the emergence of magnetic interactions.

The MR in EuZn$_{1.8}$Ga$_{0.2}$P$_2$ reflects the coexistence of multiple transport regimes governed by distinct, yet related, physical mechanisms. At high temperatures, where magnetic correlations are weak, the MR follows a semiclassical quadratic field dependence, arising from effects on charge-carrier trajectories. Upon cooling, short-range magnetic correlations develop among the Eu$^{2+}$ moments, and spin-disorder scattering becomes relevant. In this intermediate regime, the negative MR is well described by the Khosla–Fischer model, which captures the suppression of spin-disorder scattering under an applied magnetic field. At lower temperatures, the strong coupling between charge carriers and localized Eu moments favors the formation of magnetic polarons, leading to enhanced carrier localization and a pronounced negative magnetoresistance. Within this framework, the Khosla–Fischer description can be regarded as the weak-to-intermediate coupling limit of magnetic-polaron physics, while the semiclassical quadratic magnetoresistance represents the high-temperature, non-magnetic limit.

\subsection{Electron Spin Resonance}

\begin{figure}[htpb]
   \centering
   \includegraphics[scale=.34]{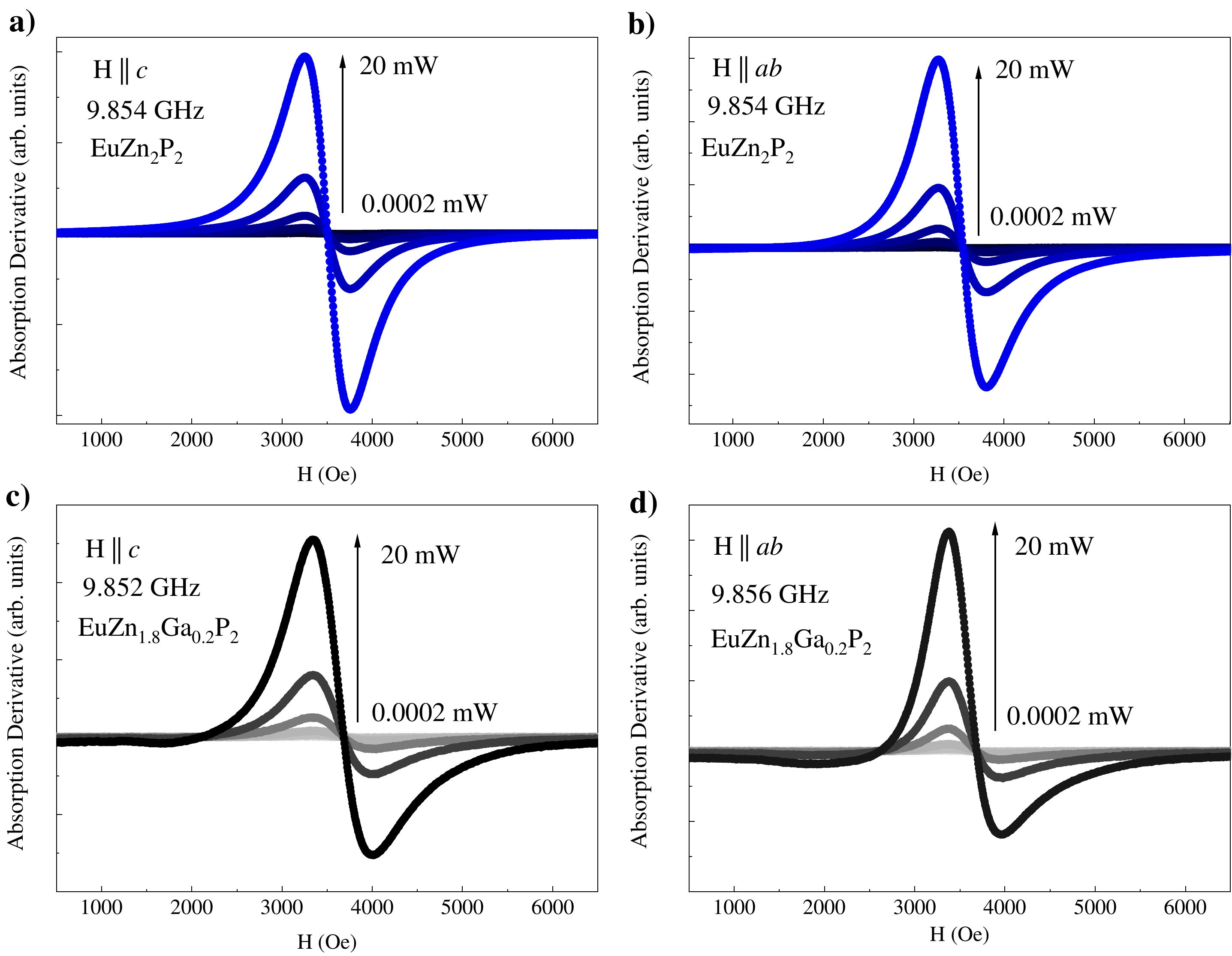}
  \caption{ESR spectra at 300~K under various microwave powers for EuZn$_2$P$_2$ for a) $H \parallel c$ and b) with $H \parallel ab$, both showing a Lorentzian lineshape. EuZn$_{1.8}$Ga$_{0.2}$P$_2$ for c) $H \parallel c$ and d) $H \parallel ab$. These spectra show a Dysonian lineshape with clear signatures of diffusion.} 
   \label{fig:EPR2}
\end{figure}

\begin{figure}[htpb]
   \centering
   \includegraphics[scale=.53]{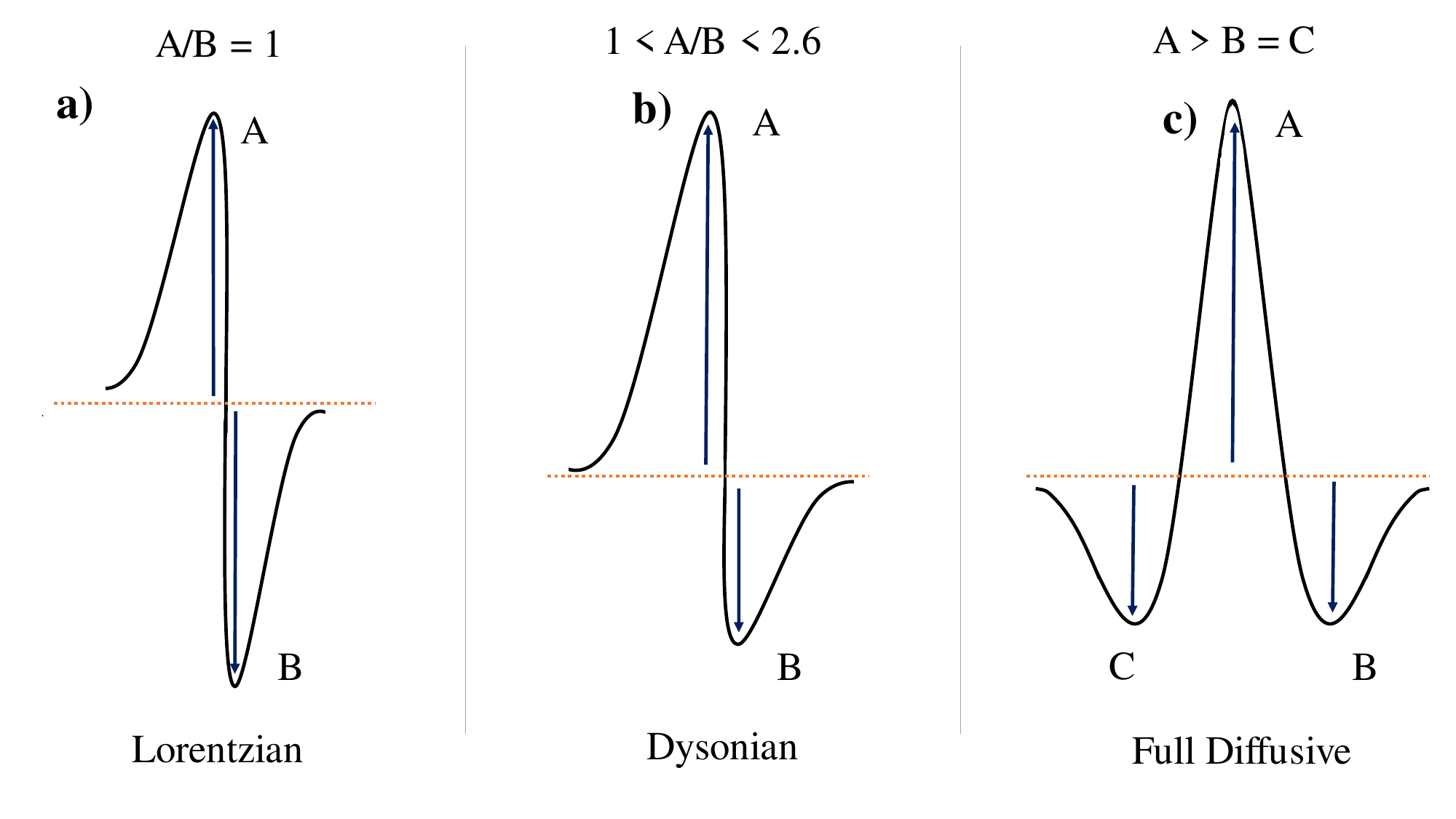}
  \caption{a) The Lorentzian lineshape is typical of insulating materials, where the microwave penetration depth is much larger than the sample size, resulting in $A/B=1$. b) The Dysonian lineshape, in contrast, is characteristic of metallic materials and arises from microwave attenuation within the sample. c) Finally, in the fully diffusive lineshape, the emergence of the $C$ dip is observed, with an amplitude comparable to $B$, as a consequence of the diffusion of conduction electrons throughout the crystal lattice.} 
   \label{fig:EPR}
\end{figure}

Electron spin resonance (ESR) spectra were measured at room temperature for EuZn$_2$P$_2$ and EuZn$_{1.8}$Ga$_{0.2}$P$_2$ single crystals with $H \parallel ab$ and $H \parallel c$, as shown in \textbf{Figure~\ref{fig:EPR2}}. As reported in previous works, EuZn$_2$P$_2$ exhibits a Lorentzian lineshape, which is characteristic of insulating materials \cite{dutra2025spin, cook2025magnetic, sichelschmidt2025electron}. This behavior arises because the microwave penetration is governed by the material’s skin depth ($\delta$): When the skin depth exceeds the crystal dimensions ($d$), skin-depth effects are negligible, because the microwaves fully penetrate the sample and are not significantly influenced by attenuation processes \cite{souza2018diffusive}. Under these conditions, a Lorentzian lineshape is expected, recognizable by the peak-to-peak amplitude ratio $A/B = 1$ (see \textbf{Figure~\ref{fig:EPR}a}). Conversely, when the skin depth is smaller than the sample size, microwave penetration is attenuated by skin-depth effects and Dysonian contributions emerge. In this regime---typical of materials with metallic behavior---the ESR signal acquires a Dysonian lineshape, for which $A > B$ and $1 < A/B < 2.6$ (see Figure~\ref{fig:EPR}b) \cite{feher1955electron,dyson1955_esr}.

\begin{figure}[htpb]
   \centering
   \includegraphics[scale=.38]{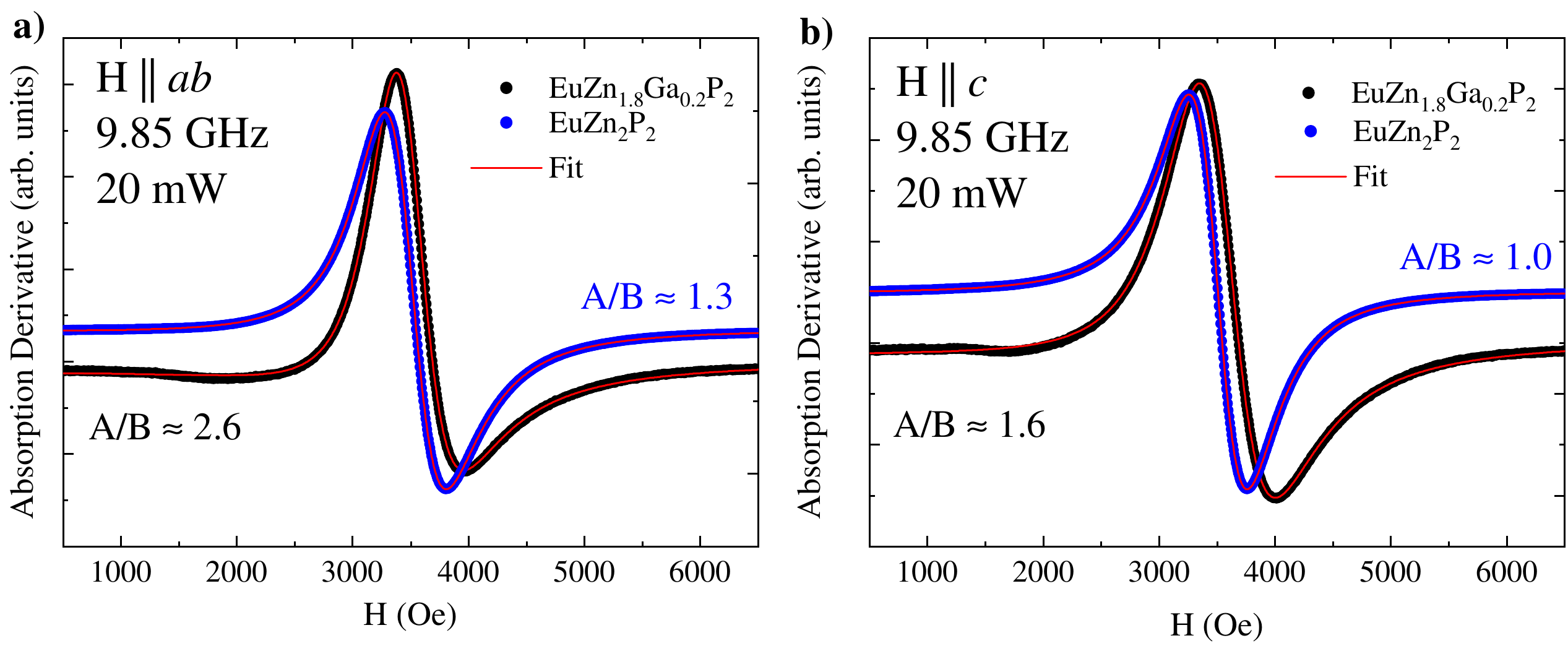}
  \caption{ESR spectra of EuZn$_2$P$_2$ (blue) and EuZn$_{1.8}$Ga$_{0.2}$P$_2$ (black) for a) $H \parallel ab$ and b) $H \parallel c$ at \qty{300}{\K}, measured with P = \qty{20}{\mW} and $\nu$ = \qty{9.5}{\GHz}. The red curve is the fit using the model presented in Equation~\ref{eq:esr}.} 
   \label{fig:EPR1}
\end{figure}

Despite EuZn$_{1.8}$Ga$_{0.2}$P$_2$ having a skin depth of $\delta = \sqrt{2\rho/2\pi\nu\mu_0}\approx$ \qty{0.5}{\mm} at \qty{300}{\K} \cite{abragam1970electron}, it displays a well-defined Dysonian shape for both field orientations, in clear contrast to EuZn$_2$P$_2$, which remains predominantly Lorentzian as shown in \textbf{Figure~\ref{fig:EPR1}}.

The only exception occurs for $H \parallel ab$, where EuZn$_2$P$_2$ shows a small asymmetry that can be captured by the Lorentzian admixture model, via absorption and dispersion coefficients ($\xi$), given by Equation~\ref{eq:esr} \cite{cabrera2015multiband}:
\begin{equation}
\frac{d[(1-\xi)\chi''+\xi\chi']}{dH} = \chi_0H_{0}\gamma_e^2T_2^2\left[\frac{2(1-\xi)x}{(1+x^2)^2}+\frac{\xi(1-x^2)}{(1+x^2)^2}\right]
\label{eq:esr}
\end{equation}
\begin{equation}
x=(H_0-H)\gamma_e T_2
\end{equation}
\noindent where $\gamma_e$ is the electron gyromagnetic ratio, $\chi_0$ is the paramagnetic contribution from static susceptibility, and $\xi$ represents the admixture of dispersion ($\xi=1$) and absorption ($\xi=0$). The Dysonian character featured in EuZn$_{1.8}$Ga$_{0.2}$P$_2$ is a classical signature of strong skin-depth effects.

\begin{figure}[htpb]
   \centering
   \includegraphics[scale=.46]{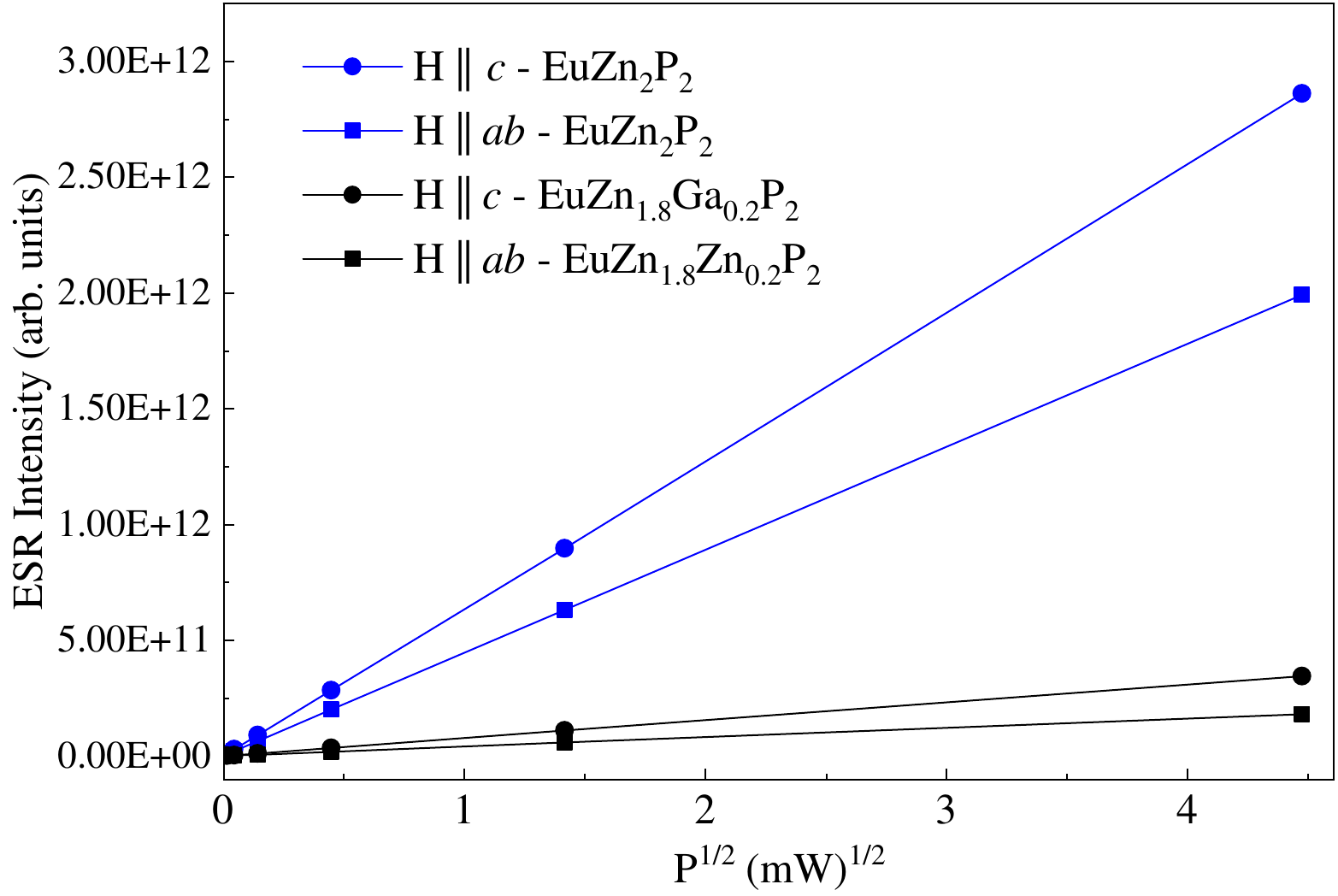}
  \caption{ESR intensity as a function of $P^{1/2}$ for EuZn$_2$P$_2$ (blue) and EuZn$_{1.8}$Ga$_{0.2}$P$_2$ (black) with $H \parallel c$ and $H \parallel ab$. Lines are a guide to the eye.} 
   \label{fig:EPR3}
\end{figure}

A marked reduction in ESR intensity occurs for EuZn$_{1.8}$Ga$_{0.2}$P$_2$ (\textbf{Figure~\ref{fig:EPR3}}). Since Ga substitution does not affect directly the Eu$^{2+}$ sublattice, a plausible explanation is a reduction in skin depth due to the increased carrier density. Enhanced microwave attenuation would reduce the effective sample volume probed, consistent with the observed lower intensity. Figure~\ref{fig:EPR3} shows the ESR intensity as a function of the square root of the microwave power. For the applied power range, the behavior is linear with no tendency toward saturation of the intensity. For all microwave powers, the ESR intensity of EuZn$_{1.8}$Ga$_{0.2}$P$_2$ remains lower than that of the pristine compound, supporting its stronger attenuation by skin-depth effects. These signatures indicate that skin-depth effects play a central role and are consistent with an enhancement of the carrier density.

Beyond the Dysonian lineshape and intensity suppression, another hallmark of conduction electrons appears through diffusive contributions to the spectrum. It is well established that both localized moments and conduction electrons can be involved in resonance transitions, although with distinct characteristics. Conduction electrons propagate across the lattice, and their dynamics can impart a diffusive component to the ESR line. When the diffusion time becomes comparable to the spin--spin relaxation time $T_2$, a fully diffusive line can emerge, characterized by the amplitude relation $A > B = C$ (see Figure~\ref{fig:EPR}c).
The presence of a diffusive contribution is inferred qualitatively from the observed lineshape asymmetry, in agreement with previous ESR studies \cite{souza2018diffusive,lesseux2016unusual}, although a fully diffusive regime is not reached in the present case. In some cases, the ESR lineshape is found to be microwave-power dependent, a behavior that has been reported in several previous studies \cite{souza2018diffusive,lesseux2016unusual}. At room temperature, however, there are no significant changes in the spectral parameters or in the lineshape, apart from the expected variation of the ESR intensity as a function of microwave power.

From the ESR data we extracted relevant parameters such as the linewidth $\Delta H$, resonance field ($H_0$), $g$-value, and $T_2$. These quantities are strongly affected by the spin--lattice ($T_1$) and spin--spin ($T_2$) relaxations, inhomogeneities, crystalline electric-field (CEF) interactions, and bottleneck effects. The fits of the model proposed in Equation~\ref{eq:esr} are shown in Table~1. From the extracted $g$-factor we obtain the $g$-shift, given by $\Delta g = g_{exp}-g_{theo}$, where $g_{exp}$ is the experimental value obtained from the ESR spectra and $g_{theo}$ is the theoretical value for insulating Eu$^{2+}$, taken as \num{1.993(2)} \cite{abragam1970electron}. In both compounds $\Delta g>0$, indicating ferromagnetic polarization of the spin carriers surrounding the Eu$^{2+}$ ions, as $\Delta g = J_{fs}\eta(E_F)$, where $J_{fs}$ is the exchange interaction between Eu$^{2+}$ and the spin carriers.

\begin{table}[h!]
\centering
\caption{Room-temperature ESR parameters of EuZn$_2$P$_2$ and EuZn$_{1.8}$Ga$_{0.2}$P$_2$ for two crystallographic orientations at \qty{20}{\mW}.}

% ===================== H || ab =====================
\begin{tabular}{lcc}
\toprule
\multicolumn{3}{c}{$H \parallel ab$} \\
\midrule
Parameter & EuZn$_2$P$_2$ & EuZn$_{1.8}$Ga$_{0.2}$P$_2$ \\
\midrule
$g$ & 2.0195& 2.0205  \\[2pt]
$\Delta H$ (\unit{\oersted}) &  \num{521(10)}&  \num{578(30)}\\[2pt]
$T_2$ (\unit{\s}) & \num{1.1014(9)e-3}& \num{1.124(2)e-3}\\
$\xi$     & \num{2.13(1)e-1}& \num{5.39(2)e-1}\\
$H_{\mathrm{0}}$ (\unit{\oersted})& \num{3486(1)}& \num{3485(1)}\\
\bottomrule
\end{tabular}

\vspace{0.5cm}

% ===================== H || c ======================
\begin{tabular}{lcc}
\toprule
\multicolumn{3}{c}{$H \parallel c$} \\
\midrule
Parameter & EuZn$_2$P$_2$ & EuZn$_{1.8}$Ga$_{0.2}$P$_2$ \\
\midrule
$g$ & 2.0085 & 1.9799\\[2pt]
$\Delta H$ (\unit{\oersted}) &  \num{501(12)}&  \num{660(13)}\\[2pt]
$T_2$ (\unit{\s}) & \num{1.1480(2)e-3}& \num{0.9123(8)e-3}\\
$\xi$     & \num{4.11(1)e-3}& \num{3.34(2)e-1}\\
$H_{\mathrm{0}}$ (\unit{\oersted})& \num{3505(1)} & \num{3555(2)}\\
\bottomrule
\end{tabular}

\end{table}

\subsection{Terahertz Time-Domain Spectroscopy}

\begin{figure}[htpb]
    \centering
    \includegraphics[width=0.8\textwidth]{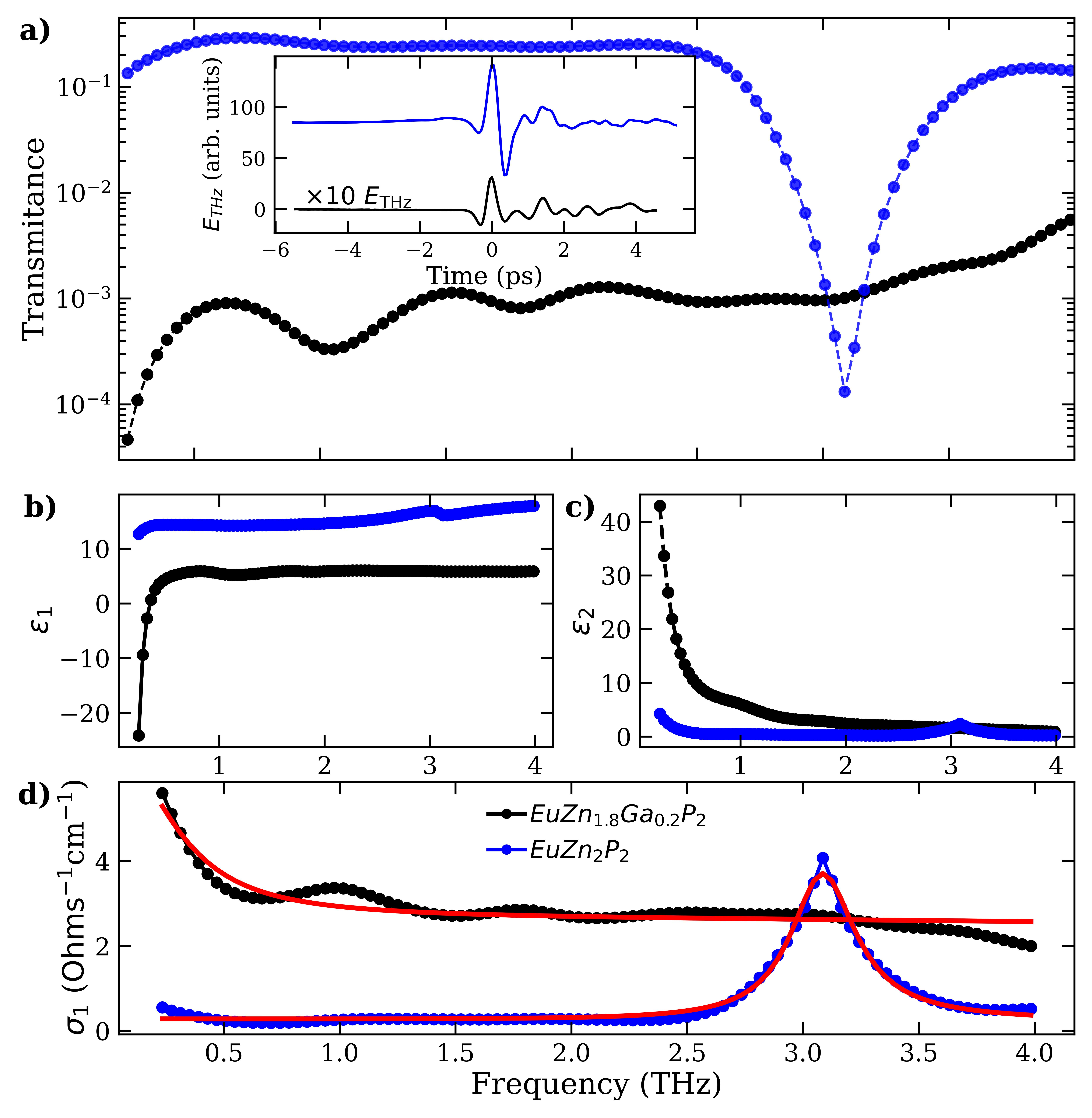}
    \caption{a) Transmittance spectrum at \qty{295}{\K} for EuZn$_2$P$_2$ (blue) and EuZn$_{1.8}$Ga$_{0.2}$P$_2$ (black), measured by terahertz time-domain spectroscopy. The inset shows the corresponding time-domain electric-field waveforms, with the EuZn$_{1.8}$Ga$_{0.2}$P$_2$ signal scaled by 10$\times$ for clarity. b) Real part ($\epsilon_1$) and c) imaginary part ($\epsilon_2$) of the experimental permittivity. d) Optical conductivity $\sigma_1(\nu)$, with solid red lines representing fits to a Drude--Lorentz model.}
    \label{figthz}
\end{figure}

We employed terahertz time-domain spectroscopy (THz-TDS) in a transmission geometry to investigate the frequency-dependent conductivity of the compounds over a broad spectral range. 
THz-TDS provides direct access to the complex transmittance  
\begin{equation}
T(\nu) = |T(\nu)|\, e^{i\phi(\nu)},
\end{equation}
where $|T(\nu)|$ and $\phi(\nu)$ denote the amplitude and phase of the transmission, respectively. The inset of \textbf{Figure~\ref{figthz}a} shows the measured time-domain waveform for each compound, from which $T(\nu)$ is extracted following the procedure described in Ref.~\cite{Marulanda2025-iy}. The main panel displays the corresponding power transmittance spectra $|T(\nu)|^2$. Compared to the pristine compound (blue), the Ga-substituted compound (black) exhibits a markedly reduced transmittance over the entire investigated frequency range, indicative of enhanced absorption induced by doping. In contrast, the undoped compound shows a pronounced resonance near \qty{3.08}{\THz}, which is strongly suppressed in the substituted compound.

The complex transmittance $T(\nu)$ directly encodes the optical response of the material in the terahertz range. For the present sample geometry, the frequency-dependent complex refractive index, $n(\nu)+i\kappa(\nu)$, can be retrieved from the measured transmittance using the conventional bulk slab approximation~\cite{neu_tutorial_2018}:
\begin{equation}
n(\nu) = 1 + \frac{c}{2\pi\nu d}\,\phi(\nu),
\label{eq:A}
\end{equation}
\begin{equation}
\kappa(\nu) = - \frac{c}{2\pi\nu d}\ln\left|\frac{\left(1+n(\nu)\right)^2}{4n(\nu)}T(\nu)\right|,
\label{eq:B}
\end{equation}
where $n$ and $\kappa$ denote the real refractive index and extinction coefficient, respectively, $c$ is the speed of light, and $d$ is the sample thickness. The complex refractive index is directly related to the dielectric response of the material through the complex permittivity, $\epsilon(\nu)=[n(\nu)+i\kappa(\nu)]^2=\epsilon_1(\nu)+i\epsilon_2(\nu)$.

Figure~\ref{figthz}b,c display the real and imaginary parts of the extracted complex permittivity for both compounds. In EuZn$_2$P$_2$, the dispersive features observed in both $\epsilon_1$ and $\epsilon_2$ near the \qty{3.08}{\THz} resonance are characteristic of a phononic excitation. This is consistent with the prediction of a phonon mode predominantly associated with Eu atomic vibrations~\cite{PhysRevB.110.014421}. At lower frequencies, the gradual increase in $\epsilon_2$ indicates the presence of free carriers, in agreement with a Drude-type response~\cite{kawahala_thickness-dependent_2023}. Notably, the Ga-substituted compound exhibits an enhanced low-frequency $\epsilon_2$, reflecting a larger Drude contribution compared to the pristine compound.

Whereas the complex permittivity describes the dielectric response of the material, expressing the data in terms of the optical conductivity, $\sigma(\nu)=\sigma_1+i\sigma_2$, enables a more direct quantitative analysis of charge transport. The experimental optical conductivity $\sigma_1$, shown in Figure~\ref{figthz}d for both compounds, is obtained from the extracted permittivity ~\cite{Lloyd-Hughes2012-oq} as the real part of:
\begin{equation}
\sigma(\nu) = -\,2\pi \nu\, i\, \epsilon_0\, \epsilon(\nu),
\end{equation}
where $\epsilon_0$ is the vacuum permittivity. The resulting spectra exhibit distinct low-frequency and resonant features, which are analyzed in the following using appropriate conductivity models.

To model the experimental spectra, the real part of the optical conductivity is expressed as the sum of contributions from free carriers and a phonon resonance~\cite{Lloyd-Hughes2012-oq},
\begin{equation}
    \sigma_1(\nu) = \sigma_\textrm{D}(\nu) + \sigma_\textrm{L}(\nu).
\end{equation}
The free-carrier response is described by the Drude term,
\begin{equation}
    \sigma_\textrm{D}(\nu) = \frac{\sigma_0}{1+(2\pi\nu\tau)^2},
\end{equation}
where $\sigma_0$ is the dc conductivity and $\tau$ is the carrier scattering time. The phononic contribution is modeled by the Lorentz term,
\begin{equation}
    \sigma_\textrm{L}(\nu) = 2\pi\epsilon_0\frac{\nu^2\gamma_\textrm{L}S_\textrm{L}^2}{(\nu_\textrm{L}^2-\nu^2)^2+(\nu\gamma_L)^2},
\end{equation}
where $S_\textrm{L}$ is the oscillator strength, $\nu_\textrm{L}$ is the resonance frequency, and $\gamma_\textrm{L}$ is the damping rate.

The solid curves overlaid in Figure~\ref{figthz}d correspond to fits of the Drude--Lorentz model to the experimental $\sigma_1(\nu)$ spectra. The model reproduces the main features of the conductivity for both compounds over the investigated frequency range. The fitting parameters extracted from this analysis are summarized in Table~2. The Eu-related phonon mode is clearly resolved in the pristine compound, yielding a resonance frequency $\nu_\textrm{L}=\qty{3.085(2)}{\THz}$ and a damping rate $\gamma_\textrm{L}=\qty{2.14(4)}{\THz}$. In contrast, in the Ga-substituted compound the phonon contribution is strongly suppressed, consistent with the dominant low-frequency absorption arising from free carriers.

\begin{table}[h!]
\centering
\caption{Fitting parameters obtained from terahertz optical conductivity}
\begin{tabular}{lcc}
\toprule
Parameter & EuZn$_2$P$_2$ & EuZn$_{1.8}$Ga$_{0.2}$P$_2$ \\
\midrule
$\sigma_0$ (\unit{\per\ohm\per\cm}) & \num{0.29(2)}  & \num{6.8(9)} \\
$\tau$ (\unit{\ps}) & \num{0.02(1)} & \num{0.4(2)} \\
\bottomrule
\end{tabular}
\end{table}

The fitting parameters indicate a dc conductivity in the Ga-substituted compound that is nearly 24 times higher than that of the pristine material. In addition, the carrier scattering time extracted for the pristine compound is consistent with the order of magnitude previously reported in Ref.~\cite{PhysRevB.108.045116} ($\sim\qty{4.6}{\fs}$). From the Drude parameters, we evaluate the ratio between the carrier density and the effective mass, commonly referred to as the Drude weight, as $N/m_\mathrm{eff}=\sigma_0/(e^2\tau)$, with $m_{\mathrm{eff}} = m^* m_0$ and $m_0$ the free-electron mass, following Ref.~\cite{stefanato2025}. This analysis yields $N/m^*=\qty{5.1(26)e16}{\per\cubic\cm}$ for the pristine, and $N/m^*=\qty{6.0(31)e16}{\per\cubic\cm}$ for the Ga-substituted compounds.

Using an effective mass of $m^*=\num{0.106}$ for both compounds, as reported for EuZn$_2$P$_2$ in Ref.~\cite{PhysRevB.108.045116}, these values correspond to estimated carrier densities of $N=\qty{5.4e15}{\per\cubic\cm}$ and $N=\qty{6.4e15}{\per\cubic\cm}$, respectively. Possible changes in the effective mass induced by Ga substitution cannot be excluded and could lead to an even larger actual difference in carrier concentrations between the two compounds. These results clearly demonstrate an enhanced carrier concentration in the Ga-substituted compound, in full agreement with the independent transport and ESR measurements presented above.

\section{Conclusions}

In summary, we have synthesized single crystals of EuZn$_2$P$_2$ and EuZn$_{1.8}$Ga$_{0.2}$P$_2$ using the Sn-flux method and demonstrated that Ga substitution produces a comprehensive reconstruction of the electronic and spin-dependent properties of EuZn$_2$P$_2$. X-ray diffraction confirms a lattice expansion upon Ga incorporation, indicating successful substitution within the [Zn$_2$P$2$]$^{2-}$ framework. 

Electrical transport measurements reveal a drastic reduction in resistivity and a suppression of the strongly activated behavior characteristic of the parent compound. The Ga-substituted compound exhibits an energy gap of $E_{\mathrm{gap}}=\qty{63}{\meV}$, a reduction of nearly \qty{45}{\percent}, together with the emergence of three distinct transport regimes linked to magnetic fluctuations. Magnetoresistance data show that colossal negative MR is preserved upon substitution, reaching $\sim\qty{90}{\percent}$ at \qty{45}{\K}. The field and temperature dependence of MR are well captured by the Khosla and Fisher semi-empirical model, indicating the continued relevance of short-range magnetic interactions and carrier scattering.

Electron spin resonance measurements at room temperature demonstrate the appearance of a pronounced Dysonian lineshape in EuZn$_{1.8}$Ga$_{0.2}$P$_2$, contrasting with the predominantly Lorentzian response of EuZn$_2$P$_2$. The Dysonian asymmetry, reduced ESR intensity, and diffusive contributions all point to a substantial enhancement in carrier density and a reduced skin depth, consistent with the electronic reconstruction induced by Ga substitution. 

Terahertz time-domain spectroscopy provides direct evidence of this enhanced free-carrier response. The Ga-substituted compound displays much stronger THz absorption and a significantly larger low-frequency optical conductivity. Drude--Lorentz modeling reveals that the dc conductivity increases by more than an order of magnitude, accompanied by a dramatic enhancement of the carrier scattering time. The extracted ratio $N/m_{\mathrm{eff}}$ further indicates an increase in carrier concentration in the Ga-substituted compound, assuming the same effective mass for both compounds. Additionally, the Eu phonon at \qty{3.085}{\THz}, clearly resolved in the pristine compound, becomes obscured by the dominant free-carrier absorption.

Overall, our results demonstrate that Ga substitution provides an effective route for tuning the electronic gap, increasing the free-carrier concentration, modifying carrier relaxation dynamics, and maintaining strong spin-dependent scattering mechanisms in EuZn$_2$P$_2$. These findings highlight EuZn$_{2-x}$Ga$_{x}$P$_2$ as a promising platform for engineering correlated narrow-gap semiconductors with enhanced transport, optical, and spintronic functionalities.

\section{Experimental Section}

\textbf{Materials}
High-purity single crystals of EuZn$_2$P$_2$ and EuZn$_{1.8}$Ga$_{0.2}$P$_2$ were synthesized by the Sn-flux method. High-purity elements, Eu (\qty{99.9}{\percent}), Zn (\qty{99.999}{\percent}), Ga (\qty{99.999}{\percent}) P (\qty{99.999}{\percent}), and Sn (\qty{99.999}{\percent}) from Alfa-Aesar were weighted in an molar ratio of 1:2:2:40 for EuZn$_2$P$_2$ and 1:1.8:0.2:2:40 for EuZn$_{1.8}$Ga$_{0.2}$P$_2$, then placed inside a quartz tube with quartz wool. The evacuated and sealed ampoule was gradually heated to \qty{500}{\celsius} over \qty{2}{\hour}, kept at this temperature for \qty{1}{\hour}, then further heated to \qty{1150}{\celsius} over \qty{4}{\hour} and held at this temperature for \qty{10}{\hour}. 
Controlled cooling was carried out down to \qty{850}{\celsius} at a rate of \qty{2}{\celsius/\hour}, after which the ampoules were rapidly spun to separate the crystals from the flux.
The samples of EuZn$_2$P$_2$ and EuZn$_{1.8}$Ga$_{0.2}$P$_2$ studied by THz-TDs had thicknesses of $d = \qty{0.213}{\mm}$ and \qty{0.155}{\mm}, respectively.

\textbf{Transport and Magnetic Characterization}
Electrical resistivity and magnetoresistance measurements was carried out using a physical properties measurements system (PPMS) in the Electrical Transport Option (ETO) with four contact probe method using Pt wires for the EuZn$_{1.8}$Ga$_{0.2}$P$_2$ samples and two contact probe method for EuZn$_2$P$_2$. The ETO works in a AC mode, and frequency was settled to \qty{70}{\Hz} and an current to \qty{0.1}{\mA} in a high impedance mode. Powder X-ray diffraction (PXRD) was carried out using a STOE STADI-P diffractometer with Cu-K$\alpha$ radiation. Electron spin resonance (ESR) measurements were performed on a Bruker-ELEXSYS 500CW spectrometer with a TE102 cavity with a modulation frequency of \qty{9.8}{\GHz} and amplitude modulation of \qty{1}{\gauss}.

\textbf{Terahertz Characterization}
Terahertz Time-Domain Spectroscopy (THz-TDS) measurements were carried out using an Er-fiber laser (\qty{80}{\MHz}, \qty{1.56}{\um}) fiber-coupled to InGaAs photoconductive antennas capable of covering the spectral range up to \qty{6}{\THz}. The THz beam is shaped and focused by four parabolic mirrors to a focal spot size of approximately \qty{1}{\mm}. Since THz radiation is strongly absorbed by water vapor, the THz-TDS measurements were performed in a dry air environment at room temperature.

\medskip
\textbf{Acknowledgments} \par %delete if not applicable))

We acknowledge support from the National Institute of Science and Technology (INCT) project Advanced Quantum Materials, funded by the Brazilian agencies CNPq (Grant No. 408766/2024-7), FAPESP, and CAPES. Financial support from CAPES, CNPq (Grant Nos. 140921/2022-2 and 88887.837417/2023-00), and FAPESP (Grant Nos. 2017/20989-8 and 2017/10581-1) is also gratefully acknowledged.
E.B.S. and J.F.O. acknowledge the Fundação Carlos Chagas Filho de Amparo à Pesquisa do Estado do Rio de Janeiro (FAPERJ) for Emeritus and PDN10 fellowships, as well as for additional financial support through Grant Nos. E-26/010.002990/2014 and E-26/210.496/2024.
F.G.G.H. acknowledges support from FAPESP (Grant Nos. 2021/12470-8, 2023/04245-0, and 2023/16742-8) and from CNPq (Grant No. 306550/2023-7). Within the USP group, E.M. acknowledges support from CAPES (Grant No. \\ 88887.007580/2024-00), and R.K.M. acknowledges support from CNPq (Grant No. 132745/2025-9).

% References
\medskip

\textbf{Conflict of Interest}
The authors declare no conflict of interest.
\medskip

\textbf{Author Contributions}
M.D., E.M., and G.G.V. generated the figures and wrote the original draft of the manuscript.
M.D., J.F.O., J.M., and E.B.S. performed the electrical resistivity and magnetoresistance experiments.
E.M. and R.K.M. performed the THz-TDS measurements.
M.D., P.C.S., R.B.D., and L.M.F. performed and analyzed the XRD data.
M.D., G.G.V., A.R.V.B., L.M.F., and J.M. analyzed and investigated the electrical resistivity and magnetoresistance data.
M.D., G.G.V., and J.M. performed and analyzed the ESR spectra.
M.A.A. supervised the project activities within the GMQ/UFABC group.
E.M. and N.M.K. analyzed the experimental THz-TDS data and contributed to its interpretation and contextualization. F.G.G.H. supervised the project activities within the GCTI-THz/USP group.
\\
\medskip

\textbf{Data Availability Statement}
The data that support the findings of this study are available from the corresponding author upon reasonable request.

% Use the following code if you wish to generate your bibliography with BibTeX;
% replace the string "MSP-template" below with the name(s) of
% the BibTeX data base(s) you want to use.
% The resulting bibliography-output (the content of the .bbl file)
% must be pasted back into this file before submission.
% Please also include your BibTeX data base file(s) in your submission
% so that we can re-run BibTeX if necessary.
%
\bibliographystyle{MSP}

\end{document}